\documentclass[aps,twocolumn,prl,superscriptaddress,letterpaper]{revtex4}

\usepackage{amsmath,amssymb}
\usepackage[dvips]{graphicx}
\usepackage{longtable}
\usepackage[usenames]{color}
\usepackage[normalem]{ulem}
\usepackage{bm}

\begin{document}

\newcommand{\B}[1]{\textcolor{blue}{#1}}
\newcommand{\R}[1]{\textcolor{red}{#1}}
\newcommand{\be}{\begin{equation}}
\newcommand{\ee}{\end{equation}}
\newcommand{\ba}{\begin{eqnarray}}
\newcommand{\ea}{\end{eqnarray}}

\title{Non-adiabatic elimination of auxiliary modes in continuous quantum measurements}
\author{Huan Yang}
\affiliation{Theoretical Astrophysics 350-17, California Institute
of Technology, Pasadena, CA 91125, USA}
\author{Haixing Miao}
\affiliation{Theoretical Astrophysics 350-17, California Institute
of Technology, Pasadena, CA 91125, USA}
\author{Yanbei Chen}
\affiliation{Theoretical Astrophysics 350-17, California Institute
of Technology, Pasadena, CA 91125, USA}

\begin{abstract}
When measuring a complex quantum system, we are often interested in only a few degrees of
freedom---the {\it plant}, while the rest of them are collected as auxiliary modes---the {\it bath}.
The bath can have finite memory (non-Markovian), and simply ignoring its dynamics,
i.e., adiabatically eliminating it, will prevent us from predicting the true quantum behavior of the plant.
We generalize the technique introduced by Strunz {\it et. al.} [Phys. Rev. Lett {\bf 82}, 1801 (1999)],
and develop a formalism that allows us to eliminate the bath non-adiabatically in continuous quantum
measurements, and obtain a non-Markovian stochastic master equation for the plant which we focus on.
We apply this formalism to three interesting examples relevant to current experiments.
\end{abstract}
\maketitle

{\it Introduction.}---Recent developments in techniques of high-precision metrology have allowed
quantum-level measurement and control of matters of at all scales, ranging from single
atoms\,\cite{coldatom_review} to macroscopic mechanical oscillators\,\cite{optomech_review}.
In these experiments, the atoms or mechanical oscillators, as objects of interest (or the {\it plant}),
are usual coupled to auxiliary degrees of freedom (or the {\it bath}), e.g. the cavity mode in cavity QED
systems, which are in turn coupled to external readout devices.  It is often desirable to obtain a
self-contained equation for the state of the plant, by eliminating bath degrees of freedom, especially
when we want to implement a real-time feedback control. In the literature, the simplest approach is to
ignore the dynamics of bath modes by assuming that they follow the plant instantaneously, and can be
adiabatically eliminated. However, this becomes inadequate when bath modes
evolve at scales longer than the plant, i.e., when the system becomes {\it non-Markovnian}.

One way to account for a non-Markovian bath is the Feynman-Vernon influence functional
method\,\cite{Feynman}. Di\'{o}si and Strunz {\it et. al.} \cite{Diosi1, Diosi2, Strunz} developed
an equivalent (but much simpler) method by {\it unraveling} the bath evolution into possible
{\it quantum trajectories}.  These trajectories are shown to drive a non-Markovian stochastic
Schr\"{o}dinger equation (SSE), which average into the exact non-Markovian master equation.
Although their model does not include measurement {\it a priori}, the SSE at the Markovian limit
can be interpreted as the evolution of the plant's pure state under continuous measurement.  In general,
however, the physical interpretation of non-Markovian SSE in terms of measurement has yet to
be clarified, as discussed by Diosi \cite{Diosi3} and Wiseman {\it et al.}\,\cite{Wiseman}.

\begin{figure}[!b]
\includegraphics[width=0.35\textwidth, bb=0 0 202 62,clip]{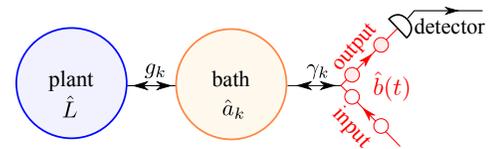}
\caption{(color online) Schematics of our measurement process. The plant is coupled to the bath, which in turn couples to an external probe field. The output probe field amplitude is projectively measured  by a detector.
\label{model}}
\end{figure}

Here we consider the non-Markovian measurement process involving a plant-bath system, in
which the bath has finite memory, and measurement is done through the bath.  We further assume
that the incoming probe field for measurement is a quantum Wiener process~\cite{Gardiner}, and
that the output field is projectively measured~\cite{comment1}. The setup is shown in Fig.\,\ref{model}.
In addition to the usual applied assumptions for the plant-bath interaction---the bath is bosonic and
couples linearly to the plant, we assume that the bath is also linearly coupled to the probe field.
We do not make assumptions on the plant, nor the plant quantity that couples to the bath.  By
generalizing the Diosi-Strunz approach, we show  that the bath can be eliminated from the full
evolution equation, resulting in a non-Markovian Stochastic Master Equation (SME) which governs
the density matrix of the plant, and thereby has a distinctive physical meaning.

The prescription we develop here can be applied to a wide range of non-Markovian quantum
measurements, thereby laying the foundation for  an interesting research direction. The purpose of this
Letter is to present the general formalism and highlight three examples of relevance to current
experiments, for which analytical forms of the SME can be obtained. Interestingly, in two of the three
cases, the non-Markovian dynamics we obtain differs from conventional wisdom.

{\it Model.}---The Hamiltonian for our model reads:
\begin{align}\nonumber
\hat H&= \hat H_p+\hat H_b+\hat H_{\rm int}+{\sum}_k\hbar \sqrt{\gamma_k}[\hat a_k\,\hat b_{\rm in} ^{\dag}(t)+\hat a_k^{\dag}\,\hat b_{\rm in}(t)],\\
\hat H_b&\equiv {\sum}_k \hbar \,\omega_k \hat a_k^{\dag}\hat a_k, \;\hat H_{\rm int}\equiv {\sum}_k \hbar \,g_k (\hat L \hat a_k^{\dag}+
\hat L^{\dag}\hat a_k).
\label{eq5}
\end{align}
Here $\hat H_p$, $\hat H_b$ and $\hat H_{\rm int}$ are the plant, bath, and interaction
Hamiltonians, respectively; $\hat a_k$ and $\omega_k$ are the annihilation operators  and
eigenfrequencies of different bath modes and $[\hat a_k,\,\hat a^{\dag}_{k'}]=\delta_{kk'}$;
the plant operator couples to the bath through $\hat L$, with $g_k$ its coupling constant to
the $k$-mode; $\hat b_{\rm in}(t)$ are annihilation operators for the input probe field at
different times and $[\hat b_{\rm in}(t),\,\hat b_{\rm in}^{\dag}(t')]=\delta(t-t')$; $\gamma_k$
is the coupling strength between the bath and the probe field. We exclude those modes that are
not coupled to the probe field, as they will simply introduce decoherence, which has already been
discussed extensively in the literature. In addition, we only consider one probe field, and can
be easily generalized to multiple probe fields.

{\it Conditional dynamics.}---At each moment, the {\it output} probe field $\hat b_{\rm out}(t)$
is projectively measured by a detector, e.g., homodyne detection if the probe field is an optical
field. We assume that (phase) quadrature $\hat b_2\equiv [\hat b_{\rm out}(t)-\hat b_{\rm out}^{\dag}(t)]/i\sqrt{2}$
is measured with the result at time $t$ being $y(t)$. Given the measurement result, the plant-bath
system is projected into a conditional state, with joint wave function $|\psi\rangle$
 at $t+{\rm d}t$ given by
\begin{align}\nonumber
|\psi(t+{\rm d} t)\rangle &= \frac{1}{P^{1/2}(y)}\langle y(t)| \hat U({\rm d}t) |{\bm 0}\rangle\otimes|\psi(t)\rangle.
\end{align}
Here $\hat U({\rm d}t)= e^{-i\hat H {\rm d}t/\hbar}$ is an evolution operator; we assume that the
{\it input} probe field (before interaction) is at vacuum state $ |{\bm 0}\rangle$ and is separable
from the joint plant-bath state; $|y(t)\rangle$ is an eigenstate of $\hat b_2(t)$; $P(y)$ is the probability
density for the measurement result and $P(y) = {\rm Tr}_{pb}\{|\psi(t+{\rm d}t)\rangle \langle \psi(t+{\rm d} t)|\}$.
By integrating over the probe field variable,  we can obtain the following nonlinear {\it Markovian}
SSE for the plant-bath state:
\begin{align}\nonumber
{\rm d}|\psi\rangle =&-\frac{i}{\hbar}(\hat H_p+\hat H_b+\hat H_{\rm int})|\psi\rangle{\rm d}t-{\sum}_{kk'}
\sqrt{\gamma_k\gamma_{k'}}\left[\hat a_k^{\dag}\hat a_{k'}\right.\\\nonumber& \left.+\langle \hat a_k-
\hat a_k^{\dag} \rangle\hat a_{k'}-\langle \hat a_k-\hat a_k^{\dag}\rangle\langle \hat a_{k'}-\hat a_{k'}^{\dag}\rangle/4
\right]|\psi\rangle {\rm d}t\\&-{\sum}_ki\sqrt{\gamma_k/2}
(2\,\hat a_k-\langle \hat a_k-\hat a_k^{\dag}\rangle)
|\psi\rangle {\rm d}W,
\label{eq1}
\end{align}
and $y(t){\rm d}t= -i{\sum}_k\sqrt{\gamma_k}\langle \hat a_k-\hat a_k^{\dag}\rangle{\rm d}t + {\rm d}W/\sqrt{2}$,
which is from the obtained measurement result distribution $P(y)={({\rm d}t/\pi)}^{1/2} \exp[-(y+i{\sum}_k\sqrt{\gamma_k}\langle \hat a_k-\hat a_k^{\dag}\rangle)^2 {\rm d}t]$ with ${\rm d}W$ being the Wiener increment (${\rm d}W^2={\rm d}t$), and
$\langle  \hat a_k\rangle\equiv \langle \psi| \hat a_k|\psi\rangle$. When the bath is a single cavity mode, it gives
the well-known Markovian SSE for the conditional evolution, also known as quantum trajectory\,\cite{Carmicheal},
of the plant and cavity mode under homodyne detection\,\cite{Wiseman2}.

{\it Elimination of bath modes.}---To non-adiabatically eliminate bath modes, we apply the method by Strunz
{\it et al.}\,\cite{Strunz} and choose unnormalized coherent-state representation $|\bm \alpha \rangle \equiv \exp
[-\sum_k \alpha_k \hat a_k^{\dag}]|{\bm 0}\rangle$ for the bath modes. One can obtain an equation for
$|\psi(\bm \alpha^*)\rangle \equiv \langle \bm \alpha |\psi\rangle $ by using
$\langle \bm \alpha | \hat a_k |\psi \rangle = \partial_{\alpha_k^*}|\psi(\bm \alpha^*)\rangle $
and $\langle \bm \alpha | \hat a_k^{\dag} |\psi \rangle = \alpha_k^* |\psi(\bm \alpha^*)\rangle $. The reduced
density matrix for the plant is given by
\be\label{eq2}
\hat \rho_p={\rm Tr}_b[|\psi \rangle \langle \psi|]=\int {\rm d}^2{\bm \alpha}\, e^{-|\bm \alpha|^2}| \psi(\bm \alpha^*)
\rangle \langle \psi (\bm \alpha) |.
\ee
By using the fact that $\int {\rm d}^2{\bm \alpha}\, \alpha_k e^{-|\bm \alpha|^2}| \psi(\bm \alpha^*) \rangle
\langle \psi (\bm \alpha) |= \int {\rm d}^2{\bm \alpha}\, e^{-|\bm \alpha|^2}\partial_{\alpha^*_k}| \psi(\bm \alpha^*)
\rangle \langle \psi (\bm \alpha) |$ and from Eq.\,\eqref{eq2},
we obtain the {\it non-Markovian SME} for the plant:
\begin{align}\nonumber
{\rm d}\hat \rho_{p}=&-\frac{i}{\hbar}[\hat H_p, \, \hat \rho_p]\,{\rm d}t-\sum_k g_k([\hat L^{\dag},\, \hat {\varrho}_k]-[\hat L,\,\hat \varrho^{\dag}_k]){\rm d}t\\&+\sum_k\sqrt{2\gamma_k}(\hat \varrho_k+\hat \varrho^{\dag}_k-{\rm Tr}_p\{\hat \varrho_k+\hat \varrho^{\dag}_k\}\hat \rho_p){\rm d}W, \label{eq6}
\end{align}
and $y(t){\rm d}t= {\sum}_k\sqrt{\gamma_k}\,{\rm Tr}_p[\hat \varrho+\hat \varrho^{\dag}]{\rm d}t + {\rm d}W/\sqrt{2}$,
where we have introduced:
\be \label{eq4}
\hat \varrho_k \equiv i\int \rm d^2{\bm \alpha}\, e^{-|\bm \alpha|^2} \partial_{\alpha_k^*}|\psi(\bm \alpha^*) \rangle \langle \psi (\bm \alpha) |.
\ee
Here the non-Markovianity only arises when we eliminate the bath, which has a memory about the plant. {\it Eqs.\,\eqref{eq6} and \eqref{eq4} will be self-contained SMEs governing the plant and measurement data, if $\hat \varrho_k$ can be written in terms of $\hat \rho_p$ and other plant operators.}
To derive $\hat \varrho_k$, we use the approach in Ref.\,\cite{Strunz} by introducing  the plant operator $\hat O_k$ as follows:
\be
\partial_{\alpha_k^*}|\psi(\bm \alpha^*) \rangle \equiv -i\, \hat O_{k}(t, \,\bm \alpha^*)|\psi(\bm \alpha^*) \rangle.
\ee
In the simplest case, $\hat O_k$ does not depend on $\bm \alpha^*$ and $\hat \varrho_k=\hat O_k(t) \hat \rho_p$.
In general, $\hat \varrho_k$ is a super-operator of $\hat \rho_p$:
\be
\hat \varrho_k=\hat {\cal A}_{0k}(t)\hat \rho_p+\hat \rho_p \hat {\cal A}_{1k}(t)+\hat {A}_{2k}(t) \hat \rho_p \hat {\cal A}_{3k}(t),
\ee
where $\hat {\cal A}_i$ are plant operators determined from $\hat O_k$.
Systematic procedures for deriving $\hat O_k$ (without measurement) has been developed by Yu {\it et al.}\,\cite{Ting}, and applied to systems with
different plant Hamiltonians. Yu's method can be generalized to our case by using interaction-picture
$|\psi(\bm \alpha^*) \rangle_I= \hat U^{-1}(t) |\psi(\bm \alpha^*) \rangle$ with a {\it non-unitary} evolution operator:
$\hat U(t)=\exp[-(i/\hbar)(\hat H_p+\hat H_b-i\hbar \sum_{kk'}\sqrt{\gamma_k\gamma_{k'}}\hat a_k^{\dag}\hat a_{k'})t]$.
We can then determine $\hat O_{k}$ from the following consistency condition \cite{Strunz, Ting}:
\be\label{eq7}
\frac{{\rm d}}{{\rm d}t}[\partial_{\alpha_k^*}|\psi(\vec{\bm \alpha}^*)\rangle_I]= \partial_{\alpha_k^*}\left[\frac{{\rm d}}{{\rm d}t}|\psi(\vec{\bm \alpha}^*)\rangle_I\right].
\ee
In general, $\hat O_k$ is difficult to solve for analytically and must be considered case by case.  In the following, we shall consider three interesting examples  that are closely related to current experiments, and analytical forms of $\hat O_k$, or equivalently $\hat \varrho_k$, can be obtained.

\begin{figure}[!b]
\includegraphics[width=0.48\textwidth, bb=0 0 347 60,clip]{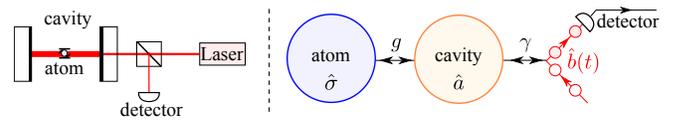}
\caption{Schematics showing the atom-cavity system.  A two-level atom (or a qubit) interacts with a cavity mode that is
coupled to an external continuous optical field which is measured via homodyne detection.
\label{atom}}
\end{figure}

{\it Atom-cavity interaction.}---As shown schematically in Fig.\,\ref{atom}, we consider the following Hamiltonian:
\begin{align}\nonumber
\hat H=&\hbar \frac{\omega_q}{2}\hat \sigma_z+\hbar \, \omega_c \hat a^{\dag}\hat a+\hbar \,g(\hat \sigma_-\hat a^{\dag}+\hat \sigma_+\hat a)+\\&\hbar\sqrt{\gamma}[\hat a\,\hat b_{\rm in}^{\dag}(t)e^{i\omega_0 t}+\hat a^{\dag}\,\hat b_{\rm in}(t)e^{-i\omega_0t}].
\end{align}
The first three terms describe the  Jaynes-Cummings-type interaction
with $\omega_q$ the atom transition frequency and $\hat \sigma_z$ the Pauli matrix, and
$\omega_c$ and $\omega_0$ are the cavity resonant frequency and the laser frequency, respectively. In the rotating frame at the laser frequency, the
Hamiltonian can be rewritten as: $
\hat H=\hbar ({\omega_q}/{2})\hat \sigma_z+\hbar \, \Delta \hat a^{\dag}\hat a+\hbar\, g(\hat \sigma_-\hat a^{\dag}+\hat \sigma_+\hat a)+\hbar\sqrt{\gamma}[\hat a\,\hat b_{\rm in}^{\dag}(t)+\hat a^{\dag}\,\hat b_{\rm in}(t)] $
with $\Delta\equiv \omega_c-\omega_0$. In comparison with the general Hamiltonian in Eq.\,\eqref{eq5}, this corresponds to the case of
$\hat L=\hat \sigma_-$ and $g_k = g\,\delta_{1k}$ (the bath has only one cavity mode and we will ignore subscript $k$ afterwards). By using the consisteny condition~\eqref{eq7}, the
operator $\hat O=f(t)\hat \sigma_-$ and $\hat \varrho$ has the following simple form:
\be
\hat \varrho = f(t)\hat \sigma_- \hat \rho.
\ee
Here the time-dependent function $f(t)$ satisfies a Riccati equation, $
\dot f-i(\omega_q-\Delta+i\gamma) f -g f^2= g\,$
with the initial condition $f(0)=0$, from the assumption that the cavity mode is initially at a vacuum state.
The corresponding SME for the atom density matrix reads:
\begin{align}\nonumber
{\rm d}\hat \rho=&-i\left[\frac{\omega_q}{2}\hat \sigma_z+g\Im\{f\}\, \hat \sigma_+\hat \sigma_-, \, \hat \rho\right]\,{\rm d}t\\\nonumber&-g\Re\{f\}\,\left[\hat \sigma_+\hat \sigma_-\hat \rho+\hat \rho\,\hat \sigma_+\hat \sigma_--2\,\hat \sigma_-\hat \rho \hat\sigma_+\right]{\rm d}t\,+\\&\sqrt{2\gamma}[f\hat \sigma_-\hat \rho+f^*\hat \rho \hat \sigma_+-\langle f\hat \sigma_-+f^*\hat \sigma_+\rangle\hat \rho]{\rm d}W.
\label{eq8}
\end{align}
This equation fully describes non-Markovian dynamics of the atom under continuous measurement. We can also obtain
the corresponding master equation if we ignore the measurement result by averaging over ${\rm d}W$ (mean of ${\rm d}W$ vanishes), namely,
\begin{align}\nonumber
\dot {\hat \rho}=&-i\left[\frac{\omega_q}{2}\hat \sigma_z+g\Im\{f\}\, \hat \sigma_+\hat \sigma_-, \, \hat \rho\right]\\&-g\Re\{f\}\,\left[\hat \sigma_+\hat \sigma_-\hat \rho+\hat \rho\,\hat \sigma_+\hat \sigma_--2\,\hat \sigma_-\hat \rho \hat\sigma_+\right].
\label{eq12}
\end{align}
This gives the exact non-Markovian master equation for a two-level atom coupled to a damped cavity mode---a dissipative environment.
Note that it differs from the one obtained by assuming a prior spectrum for the cavity mode, as have been done so far in the literature~\cite{Breuer}.

The result at the Markovian limit can be recovered by considering the case with the cavity decay rate much larger than the
atom-cavity interaction rate and also the atom transition rate, namely $\gamma\gg g$ and $\gamma\gg \omega_q$. The cavity mode memory becomes negligibly short and
\be
f(t)|_{\rm Markovian\;limit}= {g}/{\gamma},
\ee
in which case Eq.s\,\eqref{eq8} and \eqref{eq12} reduce to the usual Markovian SME and master equation, respectively.

\begin{figure}[!t]
\includegraphics[width=0.45\textwidth, bb=0 0 500 336,clip]{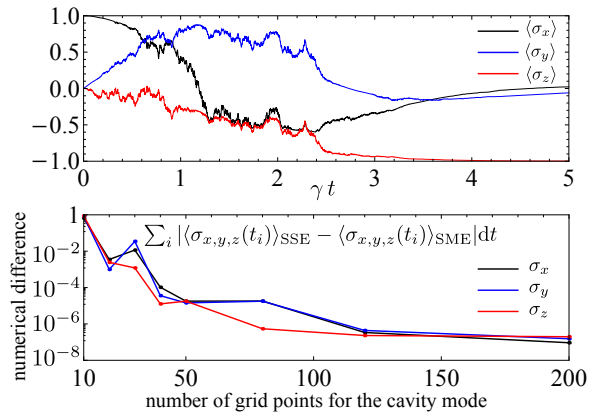}
\caption{The top panel shows numerical results of the time evolution of the conditional means:
$\langle \sigma_x \rangle$,  $\langle \sigma_y \rangle$
and  $\langle \sigma_z \rangle$  given a particular realization of ${\rm d}W$.
The bottom panel shows the convergency of the accumulated numerical difference between  the SSE and SME simulation
results given different number of grid points for the cavity mode.
\label{num}}
\end{figure}

To confirm that Eq.\,\eqref{eq8} is the SME that correctly describes
the conditional dynamics of the atom, we numerically solve (i) the Markovian SSE for the joint atom-cavity wave function and
(ii) the non-Makovian SME for the atom density matrix to see whether they both give the
same conditional mean of $\sigma_x$, $\sigma_y$ and $\sigma_z$. The numerical results are shown in Fig.\,\ref{num}. We have
chosen $\omega_q=1, \,\Delta=1$ and $\gamma=2$, and the initial state for the atom and the cavity mode is $[|+\rangle_z +|-\rangle_z]/\sqrt{2}\otimes|0\rangle$. They indeed agree with each other nicely as shown by convergency of their difference.

\begin{figure}[!b]
\includegraphics[width=0.48\textwidth, bb=0 0 360 62,clip]{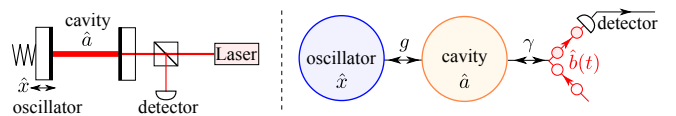}
\caption{(color online) Schematics showing a typical optomechanical device.  The mechanical oscillator is
coupled to a cavity mode via radiation pressure force. \label{optomech}}
\end{figure}

{\it Linear optomechanical interaction.}---We now consider another exactly solvable model---the linear optomechanical interaction between
a harmonic mechanical oscillator and a cavity mode. The  device is shown schematically in Fig.\,\ref{optomech}, which has been  discussed extensively in the literature recently \cite{optomech_review}. The Hamiltonian reads\,\cite{Marquardt, Rae, Genes}:
\begin{align}\nonumber
\hat H=&{\hat p^2}/({2m})+m\omega_m^2\hat x^2/2+\hbar \, \omega_c \hat a^{\dag}\hat a+\hbar\, g\,\hat x\hat a^{\dag}\hat a^{\dag}\\&+\hbar\sqrt{\gamma}[\hat a\,\hat b_{\rm in}^{\dag}(t)e^{i\omega_0 t}+\hat a^{\dag}\,\hat b_{\rm in}(t)e^{-i\omega_0t}].
\end{align}
Here $\hat x$ and $\hat p$ are the position and momentum of the oscillator with eigenfrequency $\omega_m$. Since the cavity mode usually has a large steady-state amplitude due to coherent pumping by the laser, we can consider perturbations around the steady-state amplitude and linearize the above Hamiltonian. In the rotating frame at the laser frequency, the linearized Hamiltonian is $
\hat H={\hat p^2}/{2m}+m\omega_m^2\hat x^2/2+\hbar \, \Delta \hat a^{\dag}\hat a+\hbar\, g'\,\hat x(\hat a^{\dag}+\hat a^{\dag})+\hbar\sqrt{\gamma}[\hat a\,\hat b_{\rm in}^{\dag}(t)+\hat a^{\dag}\,\hat b_{\rm in}(t)], $
where $g'\equiv g\,\bar a$ with $\bar a$ the steady-state amplitude of the cavity mode. With the same procedure as the atom-cavity case, $\hat \varrho$  can be obtained (again the bath has one mode with subscript $k$ ignored):
\be\label{eq15}
\hat \varrho= (f_1\hat \rho \hat {\cal A}^{\dag} +\hat {\cal A} \hat \rho)/(1-|f_1|^2)
\ee
with  $
\hat {\cal A}= e^{-i(\Delta-i\gamma)t}[ f_0(t) + f_x(t)\hat x] +f_p(t)\hat p$.
These functions $f_0,\,f_1,\,f_x$ and $f_p$ are determined from the consistent condition,
and satisfy coupled Riccati equations:
\begin{align}\label{eq32}
 \dot f_0=&\, i\gamma {\rm Tr}\{\hat\varrho+\hat \varrho^{\dag}\} f_1-i\sqrt{2\gamma}\,f_1\dot W -i\,\hbar g'f_0f_p\,, \\
\dot f_x=&\, e^{i(\Delta-i\gamma)t}(g'+m\omega_m^2 f_p)
-i\,g'(f_1+\hbar f_xf_p)\,,\\
\dot f_p=& -i(\Delta-i\gamma) f_p- (f_x/m)e^{-i(\Delta-i\gamma) t}-i\,\hbar g'\,f_p^2\,,\\
\dot f_1=& -i(\Delta-i\gamma)f_1+g'f_pe^{i(\Delta-i\gamma)t}-i\,\hbar \,g'f_1f_p\,.
\end{align}
These equations can be solved numerically.  Similarly, if we average the SME over ${\rm d}W$, we will
obtain the corresponding non-Markovian master equation.
It describes {\it quantum Brownian motion} of a harmonic oscillator coupled to a non-Markovian bath with dissipation, which has not yet been fully treated in the literature.

{\it Weak-coupling limit.}---In the previous cases, we took advantage of the linear interaction.
In general, when $\hat L$ is a nonlinear operator of the plant, there is
no transparent route that leads to a closed-form solution of $\hat \varrho$. If the plant-bath coupling is weak, namely $g_k<\gamma_k$,
we can perturbatively solve the problem by writing down a hierarchy of equations at different orders of $g_k/\gamma_k$.
The first-order result for the $\hat \varrho$ is very elegant:
\be\label{eq11}
\hat \varrho= \sum_{k'}\int_0^t{\rm d}\tau [e^{-i{\bf M}\tau}]_{kk'}\,g_{k'} \hat L(-\tau)\hat\rho
\ee
where $\hat L(-\tau)=e^{-i\hat H_p \tau/\hbar}\hat L\, e^{i\hat H_p \tau/\hbar}$ under free evolution.

One interesting application of this result is to study the phonon-counting experiment
recently considered in Refs.\,\cite{Martin, Thompson, Jayich, Miao}. The position of a mechanical oscillator is quadratically
coupled to a cavity mode, namely $\hat H_{\rm int}=\hbar g \hat X^2(\hat a+\hat a^{\dag})$ ($\hat X$ is the position operator
normalized by the zero-point uncertainty). If cavity bandwidth $\gamma$ is less than the mechanical frequency $\omega_m$,
only the time average of $\hat X^2$---equivalent to phonon number---is important, and we expect a direct probe of mechanical energy quantization. In the proposed experiment by Thompson {\rm et al.}\,\cite{Thompson},
the coupling strength $g$ is smaller than $\gamma$ \,\cite{comment3}; we can therefore use Eq.\,\eqref{eq11}.
From $\hat X(-\tau)=\hat X\cos\omega_m \tau-\hat P\sin\omega_m \tau$, we have
\be
\hat \varrho= \int_0^t {\rm d}\tau\, e^{-\gamma \tau} \hat X^2(-\tau) \hat \rho \approx (g/\gamma) \hat N \hat \rho
\ee
where $\hat N$ is the phonon number, and we have ignored terms proportional to $e^{-\gamma t}$, as the
characteristic measurement time scale is $t\sim \gamma^{-1}$. The resulting SME for the mechanical
oscillator density matrix reads [cf. Eq.\,\eqref{eq6}]:
\begin{align}\nonumber
{\rm d}\hat \rho=&-i[\omega_m \hat N, \, \hat \rho]\,{\rm d}t - g_{\rm eff}[\hat
X^2,\,[\hat N,\,\hat \rho]]{\rm d}t\\&+\sqrt{2g_{\rm eff}}[\{\hat N,\,\hat \rho\}-2\langle \hat N\rangle\hat \rho]{\rm d}W + {\cal O}[(g/\gamma)^2]
\end{align}
with $g_{\rm eff}=g^2/\gamma$. Note that this  does not describe a quantum non-demolition  (QND) measurement of the
phonon number, as has been argued for above, since the term $[\hat X^2,\,[\hat N,\,\hat \rho]]$ is not in the usual Lindblad form $[\hat N, [\hat N, \hat \rho]]$. It will introduce two-phonon process and cause additional diffusion; we may therefore encounter unexpected features in the actual experiment.

{\it Conclusions.}---We have reported a formalism that non-adiabatically eliminates bath modes in continuous quantum measurements and yields a self-contained non-Markovian SME for the conditional density matrix of the plant.  Conceptually, this formalism is the mathematical embodiment of how memory-induced non-Markovianity arises when we focus on a subsystem of a larger, Markovian system.  In practice, if the plant is indeed all we care about, the non-Markovian dynamics obtained here is an exact and the most efficient way of obtaining its evolution, both in terms of analytical and numerical complexity. By averaging over measurement results, the resulting master equation  describes the non-Markovian dynamics of the plant coupled to a bath that suffers from additional dissipation, a scenario not yet fully explored in the literature. We have briefly illustrated the powerfulness of this formalism using three examples, and we fully expect that it will find wide theoretical and experimental applications.

{\it Acknowledgements.}---We thank B.L.\ Hu and T.\ Yu for introducing us to this research direction and further discussions on technical details. We thank S.L.\ Danilishin and F.Ya,\ Kahlili for fruitful discussions. This work is supported by NSF grants PHY-0555406, PHY-0653653, PHY-0601459, PHY-0956189, PHY-1068881,
as well as the David and Barbara Groce startup fund at Caltech.

{\it Note added.}---During the preparation of this draft, we notice that a similar model is considered by Di\'{o}si\,\cite{Diosi4}.


\begin{thebibliography}{99}
\bibitem{comment3} The strength is the effective one for $x^2$ coupling, not the one in the strong-coupling condition shown  in Ref.\,\cite{Miao}.
\end{thebibliography}
\end{document}